\documentclass[showpacs,10pt,twocolumn,prb]{revtex4-1}
\usepackage{amsmath}
\usepackage{amssymb}
\usepackage{graphics}
\usepackage{epsfig}
\usepackage{CJK}

\begin{document}

\begin{CJK*}{GBK}{}

\title{Anisotropy of upper critical fields and thermally-activated flux flow of quenched K$_{x}$Fe$_{2-y}$Se${_2}$ single crystals}
\author{Hechang Lei and C. Petrovic}
\affiliation{Condensed Matter Physics and Materials Science Department, Brookhaven
National Laboratory, Upton, NY 11973, USA}
\date{\today}

\begin{abstract}
We report the anisotropy of the upper critical fields $\mu_{0}H_{c2}(T)$ and thermally-activated flux flow (TAFF) behavior of quenched K$_{x}$Fe$_{2-y}$Se$_{2}$. Even though the post-annealing and quenching process enhances the superconducting volume fraction, it has a minor effect on the upper critical fields for H$\parallel$c and H$\parallel$ab. Analysis of the angular-dependence of resistivity $\rho_{ab}(\theta,H)$ indicates that it follows the scaling law based on the anisotropic Ginzburg-Landau (GL) theory and the anisotropy $\Gamma(T)$ increases with decreasing temperature with $\Gamma(T)$ $\sim$ 3.6 at 27 K. The resistivity of quenched sample exhibits an Arrhenius TAFF behavior for both field directions. Field dependence of thermally activated energy $U_{0}(H)$ implies that the collective flux creep is dominant in high fields and point defects are the main pinning source in this regime.
\end{abstract}

\maketitle
\end{CJK*}

\section{Introduction}

Since the discovery of iron-based superconductors,\cite{Kamihara} these
materials have attracted great interests because of their complexity and
possible unconventional superconductivity. They contain Fe with local
moment, exhibit interplay with spin density wave (SDW) state, multiband
effects and possible s$\pm $ pairing.\cite{Cruz}$^{-}$\cite{Mazin} Among
iron-based superconductors, recently discovered A$_{x}$Fe$_{2-y}$Ch$_{2}$ (A
= K, Rb, Cs, Tl, and Ch = S, Se, Te, AFeCh-122 type) superconductors\cite%
{Guo} have some unique characteristics. These include the proximity to an
antiferromagnetic (AFM) semiconducting state distinct from SDW, possible
coexistence with long/short range AFM order, considerable Fe deficiencies in
Fe plane and the absence of hole pockets which are necessary for s$\pm $
pairing.\cite{Guo}$^{-}$\cite{Zhang Y} Structurally, AFeCh-122 materials are
rather complex, i.e, they exhibit phase separation with superconducting and
insulating regions.\cite{Wang Z}$^{-}$\cite{Li W} Insulating and
superconducting state in K$_{x}$Fe$_{2-y}$Se$_{2}$ crystals can be even
tuned reversibly by post-annealing and quenching process,\cite{Han} which is
very rare in known superconductors. It implies that superconducting and
insulating regions are intimately connected and could transform into each
other. Therefore, it is of interest to investigate the influence of
post-annealing and quenching on the superconducting properties.

In this work, we report the anisotropy of the upper critical fields $\mu
_{0}H_{c2}(T)$ and thermally-activated flux flow (TAFF) of post-annealed and
quenched K$_{x}$Fe$_{2-y}$Se$_{2}$ crystals. Our results show that the
anisotropy of $\mu _{0}H_{c2}(T)$ increases with decreasing temperature,
whereas the collective flux creep with point defects pinning source is
important at high magnetic fields.

\section{Experiment}

Crystal growth method and structure characterization of K$_{x}$Fe$_{2-y}$Se$%
_{2}$ were reported elsewhere.\cite{Lei HC2} The as-grown crystals were
sealed into Pyrex tube under vacuum ($\sim $ 10$^{-1}$ Pa). The ampoule was
annealed at 400 $^{\circ }$C for 1h and quenched in the air.\cite{Han}
Crystals were cleaved and cut into rectangular bars and the in-plane
resistivity $\rho _{ab}$(T) was measured using a four-probe configuration in
a Quantum Design PPMS-9. The sample dimensions were measured with a Nikon
SMZ-800 optical microscope with 10-$\mu $m resolution. Magnetization
measurements were performed in a Quantum Design Magnetic Property
Measurement System (MPMS-XL5).

\section{Results and Discussion}

\begin{figure}[tbp]
\centerline{\includegraphics[scale=0.42]{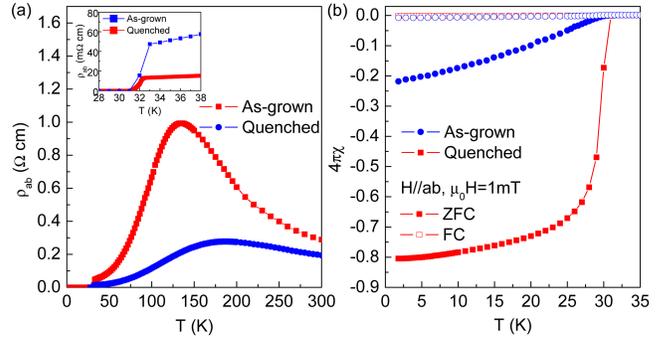}} \vspace*{-0.3cm}
\caption{(a) Temperature dependence of the in-plane resistivity $\protect%
\rho _{ab}(T)$ of as-grown and quenched K$_{x}$Fe$_{2-y}$Se$_{2}$ crystals.
Inset: enlarged resistivity curves around $T_{c}$. (b) Temperature
dependence of dc magnetic susceptibility of as-grown and quenched K$_{x}$Fe$%
_{2-y}$Se$_{2}$ crystals for $\protect\mu _{0}$H = 1 mT along the ab-plane
with zero-field-cooling and field-cooling modes.}
\end{figure}

After quenching, the resistivity $\rho _{ab}(T)$ of K$_{x}$Fe$_{2-y}$Se$_{2}$
crystals decreases significantly (Fig. 1(a)) and the crossover temperature
of metal-semiconductor transition shifts from about 134 K to 186 K. This is
consistent with the results in the literature.\cite{Han} On the other hand,
quenching has a minor effect on superconducting transition temperature\ $%
T_{c}$ which is about $\sim $ 31 K for both samples as shown in the inset of
Fig. 1(a) and (b). Although the $T_{c}$s are similar for the as-grown and
quenched samples, the quenched crystal exhibits a very sharp magnetic
transition at zero-field-cooling curve and saturates at about 10 K whereas
the diamagnetic signal increases gradually for the as-grown crystal (Fig.
1(b)). The calculated volume fraction at 1.8 K from dc magnetic
susceptibility is increased after quenching.\cite{Han} Transport and
magnetic results indicate that the post-annealing and quenching process
significantly enhances the metallicity and superconducting volume fraction
of K$_{x}$Fe$_{2-y}$Se$_{2}$ crystals.

\begin{figure}[tbp]
\centerline{\includegraphics[scale=0.42]{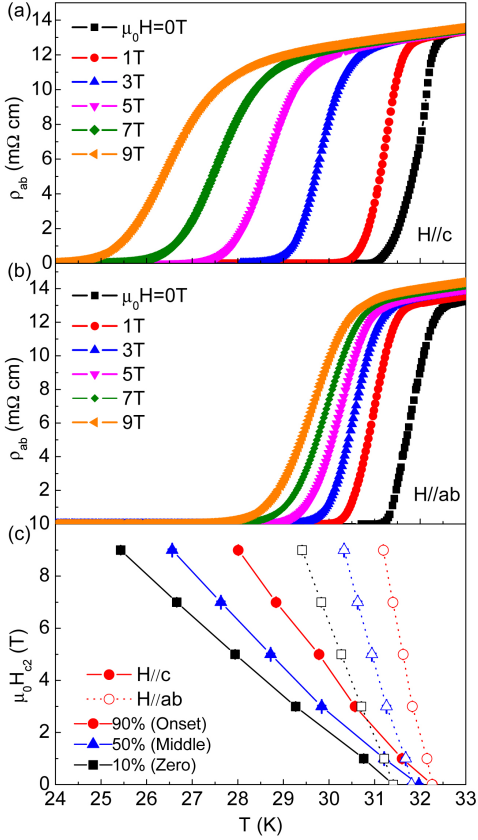}} \vspace*{-0.3cm}
\caption{Temperature dependence of $\protect\rho _{ab}(T)$ of quenched K$%
_{x} $Fe$_{2-y}$Se$_{2}$ crystal in various fields up to 9 T for (a) H$%
\parallel $c and (b) H$\parallel $ab. (c) Temperature dependence of $\protect%
\mu _{0}H_{c2}(T)$ for quenched K$_{x}$Fe$_{2-y}$Se$_{2}$ crystal determined
from the resistivity drops to 90\%, 50\%, and 10\% of the normal-state
resistivity $\protect\rho _{n}(T,H)$. The $\protect\rho _{n}(T,H)$ were
determined from the first points where the resistivity curves deviate from
the linear extrapolation of the normal-state behavior. }
\end{figure}

With increasing magnetic fields, the $T_{c}$ shifts to lower temperature for
both H$\mathbf{\parallel }$c and H$\mathbf{\parallel }$ab as shown in Fig. 2
(a) and (b).\ The shift is more pronounced when the field is parallel to the
c axis of quenched K$_{x}$Fe$_{2-y}$Se$_{2}$ crystal than the field is in
the ab plane, suggesting that the $\mu _{0}H_{c2}(T)$ is anisotropic. The $%
\mu _{0}H_{c2}(T)$ curves are nearly linear for both field directions and
the initial slope $d\mu _{0}H_{c2}/dT|_{T_{c}}$ for H$\parallel $ab is much
larger than for H$\mathbf{\parallel }$c (Fig. 2 (c) and Table 1). The slopes
are similar to reported values,\cite{Wang DM}$^{-}$\cite{Mun} indicating
that post-annealing and quenching does not have major effect on the
intrinsic superconducting properties of K$_{x}$Fe$_{2-y}$Se$_{2}$ crystals.

\begin{table}[tbp] \centering%
\caption{Upper critical fields and coherence lengths of quenched
K$_{x}$Fe$_{2-y}$Se$_{2}$ crystals.}%
\begin{tabular}{ccccccc}
\hline\hline
& $T_{c,onset}$ & \multicolumn{3}{c}{$-(d\mu _{0}H_{c2}/dT)_{T_{c}}$} & $\mu
_{0}H_{c2,zero}(0)$ & $\xi _{zero}(0)$ \\
& (K) & \multicolumn{3}{c}{(T/K)} & (T) & (nm) \\
&  & Onset & Middle & Zero &  &  \\ \hline
H$\parallel $c & 32.27(3) & 2.14(7) & 1.67(5) & 1.49(2) & 33.3(5) & 3.15(2)
\\
H$\parallel $ab & 32.26(3) & 8.3(3) & 6.0(2) & 4.46(6) & 100(1) & 1.05(2) \\
\hline\hline
\end{tabular}%
\label{TableKey copy(1)}%
\end{table}%

Using the Werthamer-Helfand-Hohenberg (WHH) formula $\mu _{0}H_{c2}$(0) =
-0.693$T_{c}$($d\mu _{0}H_{c2}/dT|_{T_{c}}$),\cite{Werthamer} and the slope
determined from 10\% $\rho _{n}(T,H)$ with $T_{c}$ = 32.27 K, the $\mu
_{0}H_{c2}$(0) is estimated to be 33.3(5) T and 100(1)\ T for H$\parallel $c
and H$\parallel $ab, respectively. The Pauli limiting field is $\mu
_{0}H_{p}(0)$ = 1.86$T_{c}(1+\lambda _{e-ph})^{1/2}$ where $\lambda _{e-ph}$
is electron-phonon coupling parameter.\cite{Orlando} Using the typical value
for weak-coupling BCS superconductors ($\lambda _{e-ph}$ = 0.5),\cite{Allen}
we obtain $\mu _{0}H_{p}(0)$ = 73.5 T for quenched K$_{x}$Fe$_{2-y}$Se$_{2}$
crystal. This value is larger than calculated $\mu _{0}H_{c2,c}(0)$ but
smaller than that for H$\parallel $ab. This could imply that the
electron-phonon coupling is strong, similar to PbMo$_{6}$S$_{8}$,\cite{Niu}
or that the real value for H$\parallel $ab may be influenced by gradual
setting in of the spin-paramagnetic effect in the high field limit. On the
other hand, experiments in high field indicate that the $\mu _{0}H_{c2,c}(0)$
is larger than the calculated value from WHH formula, suggesting that the
multiband effects might need to be considered as well.\cite{Mun} The
superconducting coherence length $\xi _{zero}$(0) estimated using the
Ginzburg-Landau formula $\mu _{0}H_{c2}(0)=\Phi _{0}/2\pi \xi ^{2}(0)$,
where $\Phi _{0}$ = 2.07$\times $10$^{-15}$ Wb is the flux quantum, is
listed in Table 1. Furthermore, the anisotropy of $\Gamma
(0)=H_{c2,ab}(0)/H_{c2,c}(0)$ is about 3, consistent with the previous
reports.\cite{Wang DM}$^{,}$\cite{Ying JJ}

\begin{figure}[tbp]
\centerline{\includegraphics[scale=0.42]{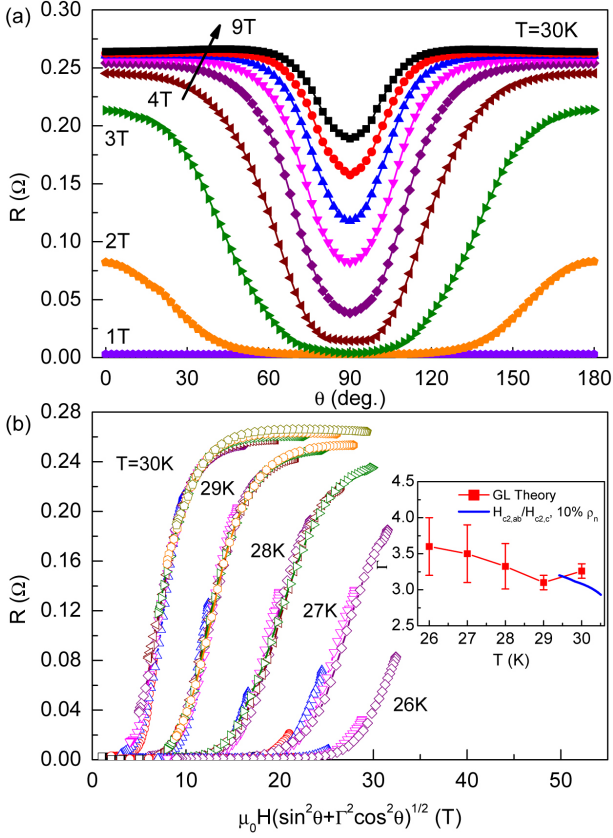}} \vspace*{-0.3cm}
\caption{Angular dependence of in-plane resistivity $\protect\rho _{ab}(%
\protect\theta ,H)$of quenched K$_{x}$Fe$_{2-y}$Se$_{2}$ crystal at 30 K in
various fields. (b) Scaling behavior of the resistivity versus $\protect\mu %
_{0}H_{s}=\protect\mu _{0}H(\cos ^{2}\protect\theta +\Gamma ^{2}\sin ^{2}%
\protect\theta )^{1/2}$ at different magnetic fields and temperatures.
Inset: the temperature dependence of determined $\Gamma (T)$ from GL theory
and $H_{c2,ab}/H_{c2,c}$ with 10\% $\protect\rho _{n}$ criterion.}
\end{figure}

Because of the uncertainty in the upper critical field values using
different criterion, there is an uncertainty in estimated anisotropy ratio $%
\Gamma (T)$. The measurements of angular-dependent resistivity $\rho
_{ab}(\theta ,H)$ can to some extent diminish this uncertainty. According to
the anisotropic Ginzburg-Landau (GL) model, the effective upper critical
field $\mu _{0}H_{c2}^{GL}$($\theta $) can be represented as\cite{Blatter}

\begin{equation}
\mu _{0}H_{c2}^{GL}(\theta )=\mu _{0}H_{c2,ab}/(\sin ^{2}\theta +\Gamma
^{2}\cos ^{2}\theta )^{1/2}
\end{equation}

where $\Gamma =H_{c2,ab}/H_{c2,c}=(m_{c}/m_{ab})^{1/2}=\xi _{ab}/\xi _{c}$.
Since the resistivity transition only depends on the effective field $%
H/H_{c2}^{GL}$($\theta $), the resistivity can be scaled using reduced $%
H/H_{c2}^{GL}$($\theta $) and all curves measured in different magnetic
fields but at the same temperature should collapse in a single curve when
choosing a proper $\Gamma (T)$ value.\cite{Blatter2} Fig. 3 (a) presents the
angular-dependent resistivity of quenched K$_{x}$Fe$_{2-y}$Se$_{2}$ crystal
at 30 K in various fields. All curves exhibit similar cup-like shape and the
minimum values of resistivity is at $\theta =90^{\circ }$, where $\theta $
is the angle between the direction of external filed and the c axis of
quenched K$_{x}$Fe$_{2-y}$Se$_{2}$ crystal. This shape indicates that the $%
\mu _{0}H_{c2,ab}$ is larger than the $\mu _{0}H_{c2,c}$, consistent with
previous results obtained when fields are fixed along ab plane or c axis.
Using scaling field $\mu _{0}H_{s}=\mu _{0}H(\sin ^{2}\theta +\Gamma
^{2}\cos ^{2}\theta )^{1/2}$ as abscissa axis and by adjusting $\Gamma (T)$,
the angular-dependent resistivity measured at same temperature in various
fields show excellent scaling behavior as shown in Fig. 3(b). Because there
is only one adjustable parameter $\Gamma (T)$ for scaling at each
temperature, the obtained value of $\Gamma (T)$ is more reliable than that
determined from $H_{c2,ab}(T)/H_{c2,c}(T)$, which may be influenced by a
choice of criterion.

As shown in the inset of Fig. 3(b), the $\Gamma (T)$ increases with
decreasing temperature and $\Gamma \ $is about 3.6 at 26 K, which is
consistent with the data on unquenched crystal.\cite{Mun} The $\Gamma (T)$
determined from $H_{c2,ab}(T)/H_{c2,c}(T)$ with 10\% $\rho _{n}$ criterion
also show same trend, but the temperature region is much higher and
narrower, which is limited by the rather high $\mu _{0}H_{c2,ab}(T)$.
Temperature dependence of $\Gamma (T)$ implies that the multiband effect may
have an influence on anisotropy of upper critical fields. The results
obtained at higher field show that the $\Gamma (T)$ decreases below 26 K,
similar to Rb$_{x}$Fe$_{2-y}$Se$_{2}$.\cite{Mun}$^{,}$\cite{Li CH} This
suggests that the spin-paramagnetic effect could also influence $\Gamma (T)$%
. The $\Gamma (T)$ of quenched K$_{x}$Fe$_{2-y}$Se$_{2}$ crystal is much
larger than of Fe(Se,Te) and Fe(Te,S).\cite{Lei HC1}$^{,}$\cite{Lei HC2x}
This is likely connected with the increase in two-dimensionality due to
enlarged inter-plane distance of FeSe tetrahedron layers by insertion of K
atoms.
\begin{figure}[tbp]
\centerline{\includegraphics[scale=0.42]{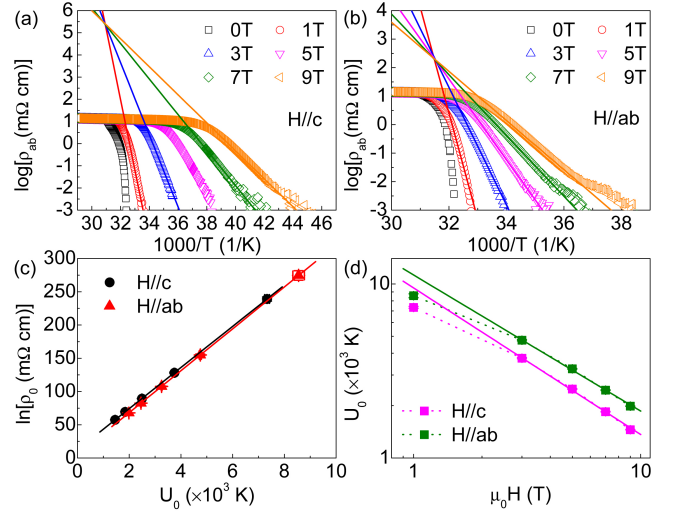}} \vspace*{-0.3cm}
\caption{$log\protect\rho (T,H)$ vs. $1/T$ in various field for (a) H$%
\parallel $c and (b) H$\parallel $ab. The corresponding solid lines are
fitting results from the Arrhenius relation. (c) $ln\protect\rho _{0}(H)$
vs. $U_{0}(H)$ derived from Arrhenius relation for both field directions.
The solid lines are linear fitting results. (d) Field dependence of $U_{0}(H)
$. The solid lines are power-law fitting using $U_{0}(H)\sim H^{-\protect%
\alpha }$.}
\end{figure}

We studied the thermally-activated flux flow (TAFF) in quenched crystal.
According to the TAFF theory, the relationship of $ln\rho $ vs. $1/T$ in
TAFF region can be expressed using Arrhenius relation,\cite{Palstra1}$^{,}$%
\cite{Palstra2}

\begin{equation}
ln\rho (T,H)=ln\rho _{0}(H)-U_{0}(H)/T
\end{equation}

where $ln\rho _{0}(H)=ln\rho _{0f}+U_{0}(H)/T_{c}$ is
temperature-independent and $U_{0}(H)$ is apparent activated energy.
Therefore, the relationship between $ln\rho (T,H)$ and $1/T$ should exhibit
linear behavior in TAFF region. As shown in Fig. 4 (a) and (b), the
experimental data can be fitted using the Arrhenius relation (solid lines)
very well for H$\parallel $c and H$\parallel $ab. The results are shown in
the common logarithmic scale in the figures but calculated in the natural
one. The good linear behavior of $ln\rho (T,H)$ vs. $1/T$ indicates that the
temperature dependence of thermally activated energy (TAE) $U(T,H)$ is
approximately linear, i.e., $U(T,H)=U_{0}(H)(1-T/T_{c})$.\cite{Palstra1}$^{,}
$\cite{Palstra2} Furthermore, as shown in Fig. 4(c), $ln\rho _{0}(H)-U_{0}(H)
$ exhibit linear behavior for both field directions which is expected from
e.q. (1). From fitting using $ln\rho _{0}(H)=ln\rho _{0f}+U_{0}(H)/T_{c}$,
we obtained $\rho _{0f}$ = 12.4(1) and 5.3(1) $\Omega \cdot cm$ and $T_{c}$
= 32.3(2) and 31.8(3) K for H$\parallel $c and H$\parallel $ab,
respectively.\ On the other hand, the $\log \rho (T,H)$ lines for different
fields extrapolate to one temperature $T_{cross}$, which should equal to $%
T_{c}$.\cite{Lei HC3} The $T_{cross}$ are about 32.2 K for both H$\parallel $%
c and 31.7 K for H$\parallel $ab, consistent with the values of $T_{c}$\
within the error bars.\ The field dependence of $U_{0}(H)$ is similar for
both field directions at high fields and can be fitted using a power law ($%
U_{0}(H)\sim H^{-\alpha }$) which is a characteristic of collective flux
creep (Fig. 4(d)).\ When $\mu _{0}H$ $>$ 3 T, the obtained $\alpha $ =
0.84(2) and 0.78(2) for H$\parallel $c and H$\parallel $ab, respectively.
Because $\alpha $ = 0.5 and 1 correspond to a planar-defect pinning and a
point-defect pinning, respectively,\cite{Chin} the fitted $\alpha $ are
close to 1, suggesting that the vortex are mainly pinned by the collective
point defects in the high field region.\ On the other hand, the experimental
$U_{0}(H)$ deviates from the extrapolated values in low field. The weaker
field dependence of $U_{0}(H)$ in low fields implies the crossover in vortex
pinning mechanism and possibly entry in the single-vortex pinning region.%
\cite{Blatter} It should be noted that the obtained $U_{0}(H)$ are much
larger than in Fe(Te,S) and comparable to a polycrystalline SmFeAsO$_{0.9}$F$%
_{0.1}$ but still much smaller than in cuprates.\cite{Lei HC3}$^{-}$\cite%
{Zhang}

Previous studies show that there is possible coexistence of superconducting
and insulating regions in K$_{x}$Fe$_{2-y}$Se$_{2}$ crystals.\cite{Wang Z}$%
^{-}$\cite{Li W} Post-annealing and quenching may enlarge the
superconducting region and/or reduce insulating region, therefore decreasing
the resistivity and improving the superconducting volume fraction of sample.
On the other hand, post-annealing and quenching has negligible effect on $%
T_{c}$, $\mu _{0}H_{c2}(T)$ and its anisotropy but it has significant
influence on the pinning force and critical current density of quenched K$%
_{x}$Fe$_{2-y}$Se$_{2}$ crystal.\cite{LEI HC4}

\section{Conclusion}

In summary, we present the superconducting properties of quenched K$_{x}$Fe$%
_{2-y}$Se$_{2}$ crystals. The results show that the post-annealing and
quenching process improves the superconducting volume fraction and reduces
scattering in K$_{x}$Fe$_{2-y}$Se$_{2}$. The hump in resistivity is shifted
to higher temperature. The $T_{c}$ and $\mu _{0}H_{c2}(T)$ of quenched K$_{x}
$Fe$_{2-y}$Se$_{2}$ crystals determined by WHH formula and anisotropic GL
theory is similar to unquenched samples, hence he quenching has minor effect
on the establishment superconducting state. The resistivity of quenched
sample shows a clear Arrhenius TAFF behavior. At high field collective flux
creep with point defects pinning center is the dominant mechanism for both
field directions whereas a possible crossover to single vortex pinning
region sets in at low field. The obtained $U_{0}(H)$ are much larger than in
Fe(Te,S) and comparable to a polycrystalline SmFeAsO$_{0.9}$F$_{0.1}$.

\section{Acknowledgements}

Work at Brookhaven is supported by the U.S. DOE under Contract No.
DE-AC02-98CH10886 and in part by the Center for Emergent Superconductivity,
an Energy Frontier Research Center funded by the U.S. DOE, Office for Basic
Energy Science.

\end{document}